\documentclass[universe,article,accept,moreauthors,pdftex,jour]{Definitions/mdpi} 

\include{journal_acronyms}
\usepackage{amssymb}

\firstpage{1} 
\pubvolume{6}
\issuenum{11}
\articlenumber{191}
\pubyear{2020}
\copyrightyear{2020}
\history{Received: 31 August 2020; Accepted: 20 October 2020; Published: 22 October 2020}
\updates{yes} 

\Title{Is OJ~287 a Single Supermassive Black Hole?}

\Author{Marina S.
 Butuzova $^{1}$*\orcidA{} and Alexander B. Pushkarev $^{1, 2}$\orcidB{}}
\AuthorNames{Marina S. Butuzova and Alexander B. Pushkarev}

\address{%
$^{1}$ \quad Crimean Astrophysical Observatory, Nauchny 298409, Crimea, Russia;  pushkarev.alexander@gmail.com\\
$^{2}$ \quad Astro Space Center of Lebedev Physical Institute, Profsoyuznaya 84/32, Moscow 117997, Russia}

\corres{Correspondence:  	mbutuzova@craocrimea.ru}

\abstract{
Light curves for more than century optical photometric observations of the 
blazar OJ~287 reveals strong flares with a quasi-period of about 12~years. 
For a long time, this period has been interpreted by processes in a binary 
black hole system. We propose an alternative explanation for this period, which is based on 
Doppler factor periodic variations of the emitting region caused by jet 
helicity. Using multi-epoch very large baseline interferometry (VLBI) observations carried out in a framework 
of the MOJAVE (Monitoring Of Jets in Active galactic nuclei with VLBA Experiments)
 program and other VLBA (Very Long Baseline Array) 
archival experiments at the observing frequency 
of 15~GHz, we derived geometrical parameters of the jet helix. To reach an 
agreement between the VLBI and photometric optical observation data, the jet 
component motion at a small angle to the radial direction is necessary. 
Such non-radial motion is observed and, together with the jet helical 
shape, can be naturally explained by the development of the Kelvin--Helmholtz 
instability in the parsec-scale outflow. In this case, the true precession of the OJ~287 jet may manifest 
itself in differences between the peak flux values of the 12-year optical 
flares. A possibility to create this precession due to Lense--Thirring effect 
of a single supermassive black hole is also discussed.}


\begin{document}

\section{Introduction}

Blazar OJ~287 is the first active galactic nucleus in the center of which it was assumed a system of two supermassive black holes.
This assumption was made by \cite{Sillanpaa88} to interpret the 100-year optical light curve of OJ~287, which shows powerful 
flares that repeat approximately every 12~years.
These flares were explained by tidal interaction of the secondary black hole with an accretion disc of the primary one at their maximum approach.
The idea of the binary black hole in the center of OJ~287 was further developed in \cite{LehtoValtonen96, Katz97, Valtonen06, Valtaoja00, ValtonenPihajoki13}, 
which suggested that the accretion disc of the primary is not in the orbital plane of the secondary component.
In these models, the orbit of the secondary black hole is precessing, thus allowing to explain the observed difference 
in time intervals between the 12-year flares, which is about a year.

The assumption about the binary black hole system in the center of OJ~287 was made before the unified scheme of the active galactic nuclei (AGN)
was proposed \cite{UP95}, which later became widely used in the interpretation of the properties of active nuclei.
According to this scheme, blazars are a class of active galactic nuclei with relativistic jets directed close to the line of sight.
The blazar OJ~287 shows strong variability in the entire observed spectral range, a high polarization degree of optical (see,~e.g.,~\cite{Takalo94} and references therein) and radio emission \citep{Cohen18}.
Therefore, we suppose that the observed optical emission is synchrotron and comes from the parsec-scale jet.
The optical emission of the accretion disc is low enough to be distinguished in the total spectrum of blazars, as it was recently proven \cite{PlavinKovPet19, KovalevZobnina20}.
Due to relativistic effects, the radiation formed in the approaching blazar jet increases in the observer's reference frame and de-boosts the emission
from the receding jet.
The relativistic boosting depends on the speed of emitting plasma and the angle of the motion with respect to the line of sight.
Thus, the periodicity in the light curve can be created by the motion of the radiating plasma along a helical trajectory, as discussed in \cite{CamKrock92}.
This scenario was applied to OJ~287 \cite{VR98}.
Namely, authors assumed the presence of two approaching relativistic jets, twisted together.
These jets are formed in a system of two black holes of the same mass.
Flares with a two-peaked profile occur at times when the first jet and later the second jet are at a very small viewing angle \cite{VR98}.
An alternative explanation is that a secondary black hole passes through the accretion disc of the primary component twice during the orbital period \cite{LehtoValtonen96}.
Taking into account the relativistic amplification of the radiation generated in the jet, we think that models explaining the 12-year optical flares by a helical jet are more~likely.

Recently, Villforth et al. (2010) \cite{Villforth10} argued the existence of a single supermassive black hole in the centre of OJ~287 based on the analysis of optical photo-polarimetric observed data for the 12-year flares. Under this assumption, the 12-year flares arose from avalanche-like accretion of the magnetic field. Britzen et al. (2018) \cite{Britzen18} had shown that the observed period of radio flux variability is about 25 years and that the changes of the inner jet position angle can be explained by the jet precession, which can be a result of either the orbital motion in the binary black hole system or the precession of the outer part of the accretion disc of the single supermassive black hole. However, there is no explanation for the difference of the variability periods observed in the radio and optical ranges.

Direct observations of parse-scale jets of active galactic nuclei can be made by very large baseline interferometry (VLBI).
For most sources, including OJ~287, VLBI maps show a typical morphology manifesting one-sided jet, the apparent origin of which is
the brightest compact feature, called the VLBI core. Due to energy losses on synchrotron radiation and adiabatic expansion, the jet quickly dims
and becomes undetectable at a distance of few milliarcseconds from the core.
As a result of the analysis of polarimetric observations of two hundred active galactic nuclei performed with the very long baseline array 
quasi-simultaneously at frequencies from 8.1 to 15.4~GHz, it was showed that the VLBI core observed at a certain frequency is a part of a jet 
in which the medium becomes optically thick for radiation at this frequency due to opacity caused by synchrotron self-absorption \cite{Sokolovsky11, Pushkarev12}.
Due to the fact that the magnetic field strength and density of radiating particles decrease with the distance from the true 
jet base according to the power-law dependence \cite{BK79, Lobanov98}, the lower the frequency of observations, the further from the true jet base
the absolute position of the VLBI core.
This implies that the observed optical radiation comes from a region close to the true jet base.
Therefore, if we associate the 12-year optical flares with a cyclic change in the angle between the velocity vector of the radiating region and 
the line of sight, we can expect similar periodicity in the data of radio observations.
A comparison of the flux density of AGN measured with VLBI and single-dish observations showed that VLBI core emission
significantly contributes and often dominates in a total source radiation \cite{Kovalev05}.
Hence, in measurements of the total radio flux density from the entire source, the periodicity cannot be strongly obscured by radiation from 
the optically thin part of the jet. Statistical analysis performed in \cite{ Ryabov16, Sukharev19} based on data from the Owens Valley Radio Observatory (OVRO,
2008$-$2018) and Radio Astronomy Observatory of Michigan University (UMRAO, 1974$-$2011) shows that the period of long-term radio variability is about 25 years. This period washed out the period of about 6 years, 
obtained previously using shorter data set \cite{Donskykh16}.
We also note that the periods obtained from the total radio flux density ($\approx$25~years \cite{ Ryabov16, Sukharev19}) and from the motion of the 
jet ridgeline ($\approx$22~years \cite{Britzen18}) are consistent.
The periodicity of $\approx$1 year is found in both optical and radio domains \cite{Britzen18, Sukharev19}.
We do not consider this short periodicity here, though it might reflect the rotational motion of the plasma in the jet flow \cite{But20} or 
could be caused by nutation-like motion in the precessing jet \cite{Katz97, Britzen18}.
We believe that the issue of the absence of a prominent variability in the radio range with a period clearly seen in the optical domain is fundamental 
for understanding both the processes operating in OJ~287 and for determining the properties of the central object.

The layout of this paper is the following. 
In Section~2, we derive geometrical and kinematic parameters of the OJ~287 jet assuming that it has a helical shape and using
15~GHz VLBA observations since 1994 through 2019, performed in the framework of the Monitoring Of Jets in Active galactic nuclei with VLBA Experiments (MOJAVE\footnote{ http://www.physics.purdue.edu/astro/MOJAVE/index.html}) project; we also provide interpretation of the difference between the variability periods in the radio and optical ranges.
In Section~3, we assume that different values of the maximum flux density of the 12-year flares can be caused by precession of the helical jet.
We showed that the period of this precession is consistent with the Lense--Thirring precession of a single supermassive black hole or the 
inner part of its accretion disc. Discussion of the obtained results and conclusions are presented in Section~4.

\section{Helical Jet Model}

In this section, we introduce geometrical parameters used by us to describe a jet helical shape and motion of jet's components.
Then, we estimate these parameters from the VLBI observational data.
In the framework of the obtained geometrical and kinematic model, we explain the observed difference in periods of flux variability in the radio and optical ranges.

\subsection{Description of Parameters}

\begin{figure}
\centering
\includegraphics[width=12 cm]{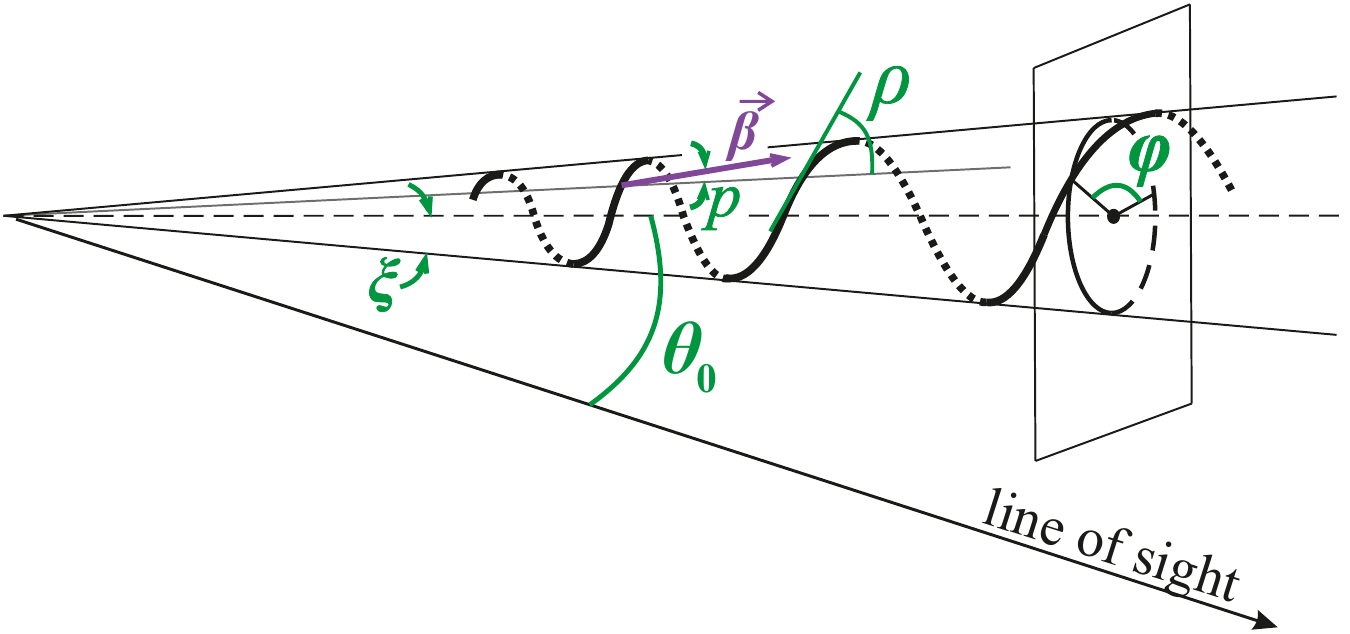}
\caption{
Scheme of a helical jet with the specified model parameters:
The cone axis is shown by the dashed line.
The thin solid line displays the cone generatrix.
The thick line denotes the jet.
The dotted line refers to those parts of the jet which are located opposite to the observer side of the cone. 
}
\label{fig:heljetscheme}
\end{figure}

Here, we consider the jet as a continuous flow, having a helical shape. According to \cite{Konigl81, Pushkarev12}, magnetic field strength and number density of emitting particles decrease with the distance from the jet beginning. Due to synchrotron opacity, the emission at different frequencies arises at different parts of this continuous jet.  
We used the geometrical description of a helical jet proposed in \cite{But18a}.
Namely, we assume that the jet axis forms a helix located on the surface of a notional cone with a half opening angle $\xi$ (Figure~\ref{fig:heljetscheme}).
The angle of the cone axis with the line of sight is $\theta_0$.
Note that the cone apex is not necessarily expected to coincide with the supermassive black hole location.
The jet swirl is characterized by the parameter $\rho$, which is the angle between the tangent to some point of the jet axis and the cone generatrix drawn through this point.
To describe the jet kinematics, the jet was divided into components by cross-sectional drawing such that the jet axis can be considered a straight line within each component.
Then, $\beta$ is a speed of the jet component and $p$ is an angle between the velocity vector and a cone generatrix passing through the component.
The jet component location at the cone surface relative to the observer is characterized by an azimuth angle $\varphi$.
The origin for $\varphi$ is located on the surface farthest from the observer cone and lies in the plane containing the line of sight and the cone axis.
The angle $\varphi$ is measured counter-clockwise.
We assume that the considered parameters do not change with time and that they are the same for all components located within the considered jet, which is up to a dozen parsecs in the projection on the sky plane.
Note that we do not associate the jet components directly with the jet features observed with the VLBA, as their physical nature 
may be different, but these features reflect the position of the jet flow on the sky plane.

\subsection{Estimation of Parameters by VLBI Data}

A difference of the jet from straight flow appears in changes of jet feature positions onto the plane of the sky.
A jet feature position projected onto the sky is characterized by a position angle (PA) measured toward the rise of right ascension  
between directions from the VLBI core to the north and the given~feature.

\begin{figure}[H]
\centering
\includegraphics[width=15 cm]{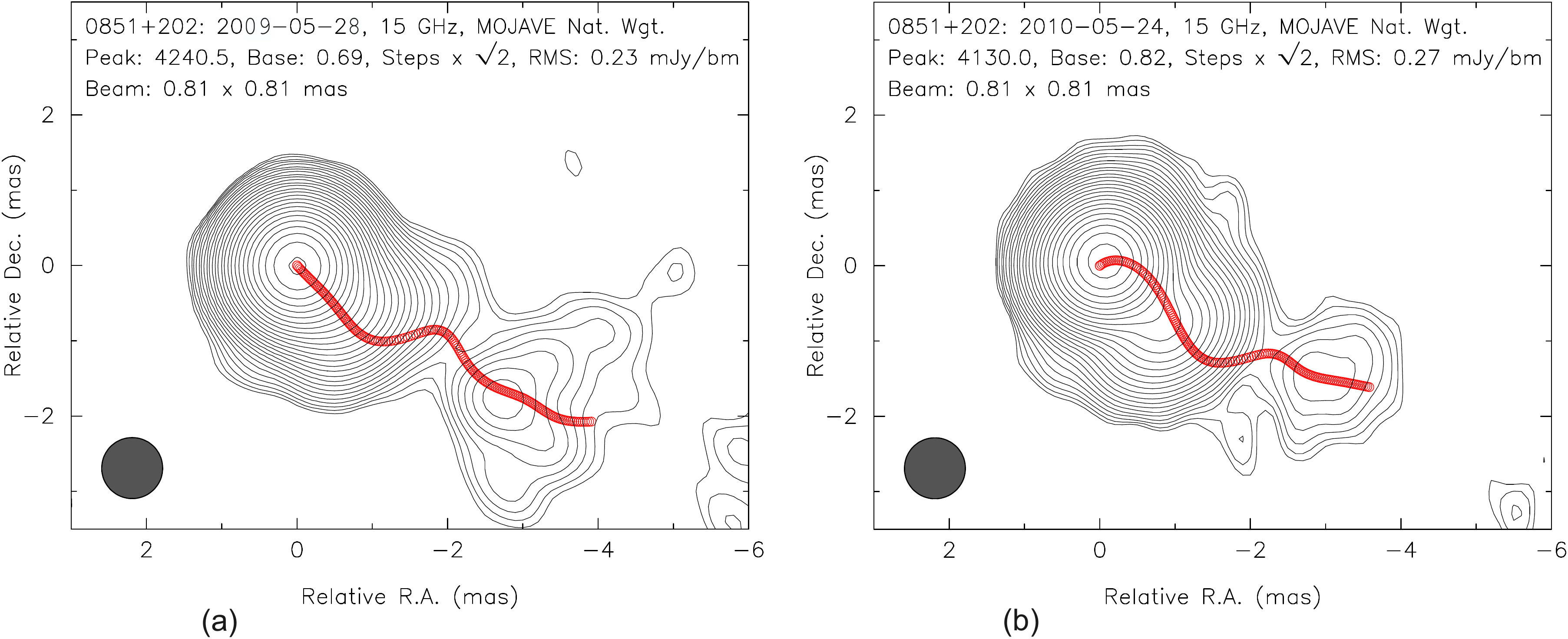}
\caption{
Contour images of the source with ridgeline overlaid at two epochs, before \textbf{(a)} and after \textbf{(b)} the abrupt change in the inner jet position angle.}
\label{fig:contourRL}
\end{figure}

\begin{figure}[H]
\centering
\includegraphics[width=10 cm]{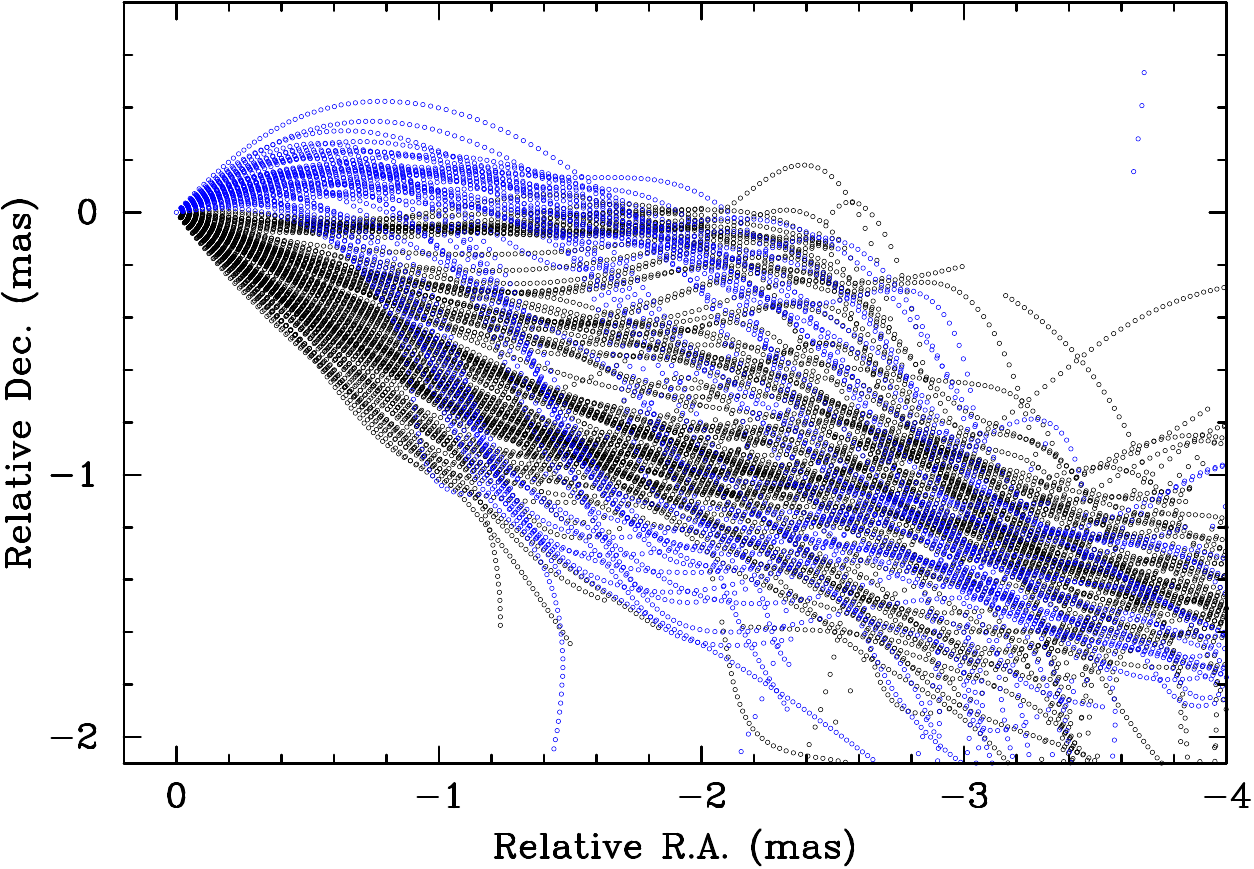}
\caption{
Total intensity ridgelines of OJ~287 for 145 epochs of Very Long Baseline Array (VLBA) observations carried out at 15~GHz within a period 1994--2019:
The origin of coordinates corresponds to the position of the 15~GHz very large baseline interferometry (VLBI) core.
}
\label{fig:ridgelines}
\end{figure}

We made use of a set of 145 epochs of VLBA observations of OJ~287 carried out at 15~GHz since 1994 through 2019.
Most of the data are taken within a framework of the ongoing MOJAVE program, while the rest are from the VLBA achival
experiments.
The core component was identified based on results of structure model fitting as an apparent origin of the jet.
For each epoch, we constructed a ridgeline of the jet in total intensity using fully calibrated data
in the image plane (Figure~\ref{fig:contourRL}). Not to be affected by the elliptical restoring beam, which differs from session to session,
we convolved the images with the same circular beam, averaged over all epochs. The algorithm for ridgelines starts 
with making azimuthal slices at progressive separations (with a step of 0.1 mas) from the VLBI core position and
finding the points where the integrated intensity across the outflow is equal on both sides of the arc. A cubic spline 
is then fit to the points and interpolated. The obtained ridgelines are presented in Figure~\ref{fig:ridgelines}.
First, the ridgelines are bent.
An apparent bend angle is amplified by the projection effects due to a small jet viewing angle. 
Second, ridgelines at different epochs have noticeable shifts from each other.
Third, the direction of the inner jet ridgelines (e.g., within 0.2 mas from the VLBI core) significantly changes, with the magnitude of $\alpha_\text{obs}\approx90^\circ$ during the first half of 2010. 
The ridgelines after the rapid change in PA of the inner jet direction (since May 24, 2010) are shown by blue color.

The innermost jet direction can be derived even more accurately by modelfitting the restored brightness distribution performed in the visibility domain.
For this, we used the procedure {\it ``modelfit"} in the Difmap package \cite{difmap}. 
The source structure was modeled with an elliptical and circular Gaussian components for the core and the inner jet feature, respectively. 
Thus, the position angle of the major axis of the core component gives the direction of the innermost jet. 
This approach provides the highest effective angular resolution as the core is the brightest feature having the highest dynamic range of the restored source structure.

\begin{figure}
\centering
\includegraphics[width=14 cm]{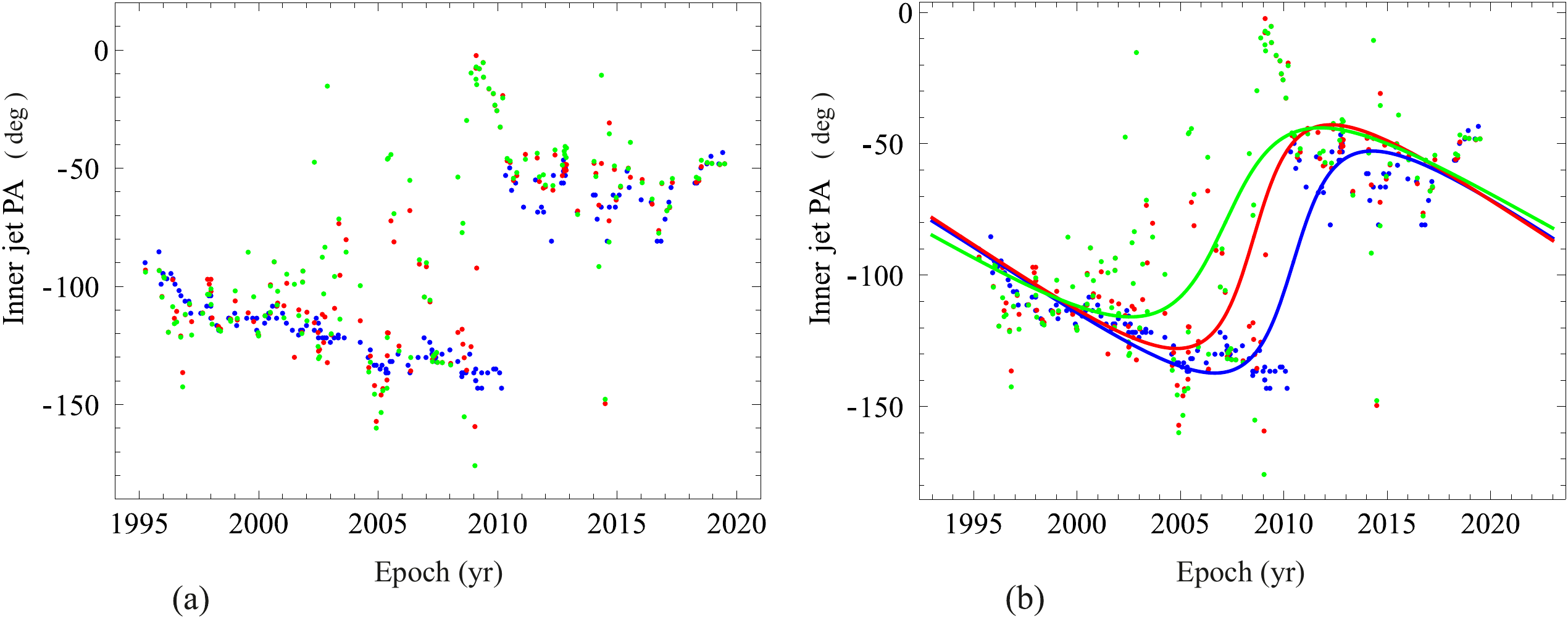}
\caption{ \textbf{(a)} Evolution of PA$_\text{in}$ of the innermost jet direction in OJ~287.
Blue points correspond to the PA$_\text{in}$ estimates derived along the jet ridgelines within $0.025$~mas from the 15~GHz VLBI core.
Red and green points mark PA$_\text{in}$ obtained from major axis orientations of the core fitted with an elliptical Gaussian component 
in addition to one or two circular Gaussian components fitted to the rest of the jet, respectively.
\textbf{(b)} Approximation for the observed $\text{PA}_\text{in}$ under the helical jet model. 
The colors of filled lines correspond to those of the observed points.
}
\label{fig:PAin}
\end{figure}

We plotted changes of the innermost jet position angle ($\text{PA}_\text{in}$) with time in three ways (Figure~\ref{fig:PAin}).
In the first way, the PA is calculated for the position of the ridgeline point closest to the 15~GHz core feature.
In the other two ways, the VLBI core was fitted by an elliptical Gaussian component.
In the second way, the fitting of the jet was performed with one circular Gaussian component, while in the third way, the fitting was performerd with two circular Gaussian components,
in addition to the elliptical Gaussian for the core. The third approach models the rest of the jet better and could be considered the most accurate way of assessing PA$_\text{in}$ values.
In the last two ways, the PA of major axis of the core fitted component was taken as $\text{PA}_\text{in}$.
The median size of the major axis of the core component is 0.19~mas, suggesting that we are sensitive to the innermost jet direction at core separations $\le0.1$~mas.

Figure~\ref{fig:PAin} shows that, despite the available spread of points defined in various ways, there is a general trend of decreasing $\text{PA}_\text{in}$ until 2009--2010, followed by an abrupt increase of $\text{PA}_\text{in}$ by about $100^\circ$ and its further gradual decrease.
Note that the sudden change of $\text{PA}_\text{in}$ defined by the major axis of the elliptical Gaussian component fitted VLBI core occurs about a year earlier than the sudden change of $\text{PA}_\text{in}$ defined by the innermost part of the jet ridgeline.
This suggests that the cause or conditions in the jet which result in the sudden change in $\text{PA}_\text{in}$ values propagate downstream the jet.
Noteworthily, $\text{PA}_\text{in}$ values obtained from 43~GHz VLBA observations correspond to those measured at 15~GHz, but the rapid change in the 43~GHz $\text{PA}_\text{in}$ occurred in the middle of 2004 (Figure~6  in \cite{Cohen17}).
This can be explained by weaker opacity of synchrotron emission at 43~GHz, which is generated in the jet region located upstream from the 15~GHz VLBI core.

Cohen et al. \cite{Cohen17} interpret the abrupt change in PA$_\text{in}$ as a shift of the jet nozzle in a new direction.
For this scenario though, it is problematic to explain the presence of a trend of decrease of PA$_\text{in}$ with time after the PA$_\text{in}$ sudden change, the slope of which is similar to that before the PA$_\text{in}$ rapid rotation.
On the other hand, the helical jet considered here can explain the observed change in PA$_\text{in}$ provided that the ratio $\theta_0/\xi$ is slightly more than unity \cite{But18a}. 
According to \cite{But18a}, the change in PA$_\text{in}$ as a function of the azimuth angle $\varphi$ is given as  
\begin{equation}
    \text{PA}_\text{in}\left(\varphi \right) \approx - \arctan \left(\frac{\sin \varphi}{\theta_0/\xi+\cos \varphi} \right)+\text{PA}_0,
    \label{eq:PAvariation}
\end{equation}
where PA$_0$ is a mean position angle (PA of the cone axis). There is a minus sign before the arc tangent because, unlike the considered case in \cite{But18a}, a smooth decrease in PA$_\text{in}$ occurs before and after the sudden change.
That is, in our case, the jet helix twists in the other direction: clockwise.
We fitted PA$_\text{in}$ changes using Equation~(\ref{eq:PAvariation}) and expressing the azimuth angle as $\varphi=2\pi t/T+\varphi_0$, where $T$ is the period of PA$_\text{in}$ change, $t$ is the time in the observer's reference frame, and $\varphi_0$ is the initial phase.  
The curves fitted to the observed change in PA$_\text{in}$ are shown in Figure~\ref{fig:PAin} (right panel), and the obtained parameters for the fits are listed in Table~\ref{tab:PAfits}. 
There is a good agreement between the values of $T$ and $\theta_0/\xi$ obtained from the three different ways of determining of PA$_\text{in}$.

\begin{table}[H]
\caption{Parameters of the helical jet derived from an approximation of the observed PA$_\text{in}$.}
\label{tab:PAfits}
\centering
\begin{tabular}{lccc}
\toprule
\textbf{Parameter}	& \textbf{PA (ridgeline)}	& \textbf{PA (core $+$ 1 jet component)} & \textbf{PA (core $+$ 2 jet component)} \\
\midrule
$\theta_0/\xi$		& $1.49\pm0.04$			& $1.48\pm 0.08$  & $1.67\pm 0.12$ \\
$T$ (yr)		& $28.7\pm1.1$			& $28.3 \pm 2.0$  & $30.7\pm 3.2$\\
PA$_0$ (deg)   & $-95.0\pm2.4$     & $-85.4 \pm 2.3$ & $-80.0\pm 2.7$ \\
$\varphi_0$ (deg)  & $348.0 \pm 6.5$   &   $10.8 \pm 11.6$ &   $31.4 \pm 15.6$ \\
goodness of fit: &   &   &   \\
$R^2$   &   0.987   &   0.934   &   0.893   \\
adjusted $R^2$  &   0.986   &   0.932   &   0.890   \\
\bottomrule
\end{tabular}
\end{table}

From Figure~\ref{fig:PAin} and Table~\ref{tab:PAfits}, it is seen that the approximations of PA$_\text{in}$ values, which are found from the inner jet ridge lines and from the source structure modeled by two components (an elliptical Gaussian for the VLBI core and a circle Gaussian for the jet), are similar.
The difference between them in $\varphi_0$ and PA$_0$ can be explained by the fact that, in the second case, PA$_\text{in}$ is measured by radiation detected from smaller angular scales than in the first case.
Additionally, in the second case, the PA$_\text{in}$ is sensitive to unresolved bright jet parts located upstream of the region where the PA$_\text{in}$ is measured from the ridgeline.
The qualitative difference between the approximation of PA$_\text{in}$ obtained from PA of an elliptical Gaussian component of the VLBI core under the jet fitting by two circular Gaussian components and the other two approximations is probably caused by a large spread of values. 

For further calculations, we use the values $T_\text{PA}=28.3$~years and $\theta_0/\xi=1.5^\circ$.
Analysis of the VLBA images stacked over all available epochs at 15~GHz for a given source for a sample of a few hundred AGN showed that jet features propagate within a certain cone \cite{Pushkarev17}.
This result is  consistent with the assumption considered by us about the jet shape and allows to accept $\xi=1^\circ$ to further calculations.
Then, $\theta_0=1.5^\circ$, which agrees with other estimations of the jet angle with the line of sight performed by the combined analysis of the jet kinematics and the long-term radio variability \cite{Hovatta09, Pushkarev09}.
The helical jet parameters found by us are listed in Table~\ref{tab:heljetpar}. 

\begin{table}[H]
\caption{Geometric and kinematic parameters derived for the blazar OJ~287 helical jet on parsec scales.}
\label{tab:heljetpar}
\centering
\begin{tabular}{lcc}
\toprule
\textbf{Parameter}	& \textbf{Sign}	& \textbf{Value} \\
\midrule
half-opening angle of the cone		& $\xi$			& $1^\circ$  \\
angle between the cone axis and the line of sight & $\theta_0$ & $1.5^\circ$    \\
variability period of PA$_\text{in}$ & $T_\text{PA}$    & 28.3~years    \\
speed of the jet component  & $\beta$   &   0.9979  \\
angle between the jet velocity vector and a radial trajectory & $p$ &   $2.5^\circ$ \\
angle between a tangent to the jet flow and a cone generatrix at a given point  &   $\rho$  & $10^\circ$    \\
\bottomrule
\end{tabular}
\end{table}

We emphasize the following fact.
The obtained variability period of PA$_\text{in}$, caused by a cyclic change of the azimuth angle of the helical jet components located at the fixed distance from the cone apex, is consistent with the variability period of the radio emission at cm wavelengths \cite{Ryabov16, Britzen18, Sukharev19}.
It can mean that the long-term radio variability is also connected with the jet helical shape causing the periodical variations of the jet angle with the line of sight.
As we mentioned above, this scenario seems to be the most probable for the interpretation of the 12-year optical flares as well.
The explanation of the difference in the periods of long-term radio and optical flux variability is given in the next section.

\subsection{Connection between Long-Term Optical and Radio Flux Variability}

When the emitting region moves with an ultrarelativistic velocity $\beta$ at a small angle $\theta$ to the line of sight, the spectral flux density of radiation formed in this region increases in the observer's reference frame
\begin{equation}
F\left( \nu \right)=\delta^{2+\alpha}F'\left(\nu \right),
    \label{eq:flux}
\end{equation}
where $\alpha$ is a spectral index of radiation described by a power-law $\left( F\propto \nu^{-\alpha}\right)$.
The Doppler factor is
\begin{equation}
 \delta=\frac{\sqrt{1-\beta^2}}{1-\beta\cos \theta}.   
    \label{eq:doppler}
\end{equation}
We estimate the component speed from the apparent speed of the jet features $\beta_\text{app}$.
Kutkin et al. \cite{Kutkin19} showed that the jet speed, which is found by time delay of a flare when the observed frequency decreases, corresponds to the maximum $\beta_\text{app}$ estimated from the apparent motion of jet components. 
Therefore, we used the maximum value of $\beta_\text{app}=15.31$ for OJ~287 \cite{Lister19}  to find  $\beta=0.9979$ assuming $\Gamma_\text{min}=(1+\beta_\text{app}^{2})^{1/2}$.

If $\beta=const$, changes in $\delta$ occur due to changes in $\theta=\theta\left( \varphi \right)$.
In order for the changes in $\delta$ to be reflected in the observed flux variability, the radiating region must be compact enough.
That is, the angle $\varphi$ should change slightly within this region.
Optical emission comes from near to the true jet base region \cite{BK79}.
This region is characterized by high magnetic fields and accelerated particles that have not yet lost a greater part of their energy through radiation.
The magnetic field strength and number density of emitting particles decrease according to a power-law with distance from the true jet base \cite{BK79, Lobanov98,Pushkarev12}.
These gradients together with high radiation energy losses of ultrarelativistic electrons via synchrotron optical emission allow us to assert that the region in which the observed optical radiation is generated is compact.
Based on the Blandford-K{\"o}nigl jet model \cite{BK79}, for further calculations, we assume that the optical emitting region is located at a fixed distance from the true jet base. 
Therefore, we can expect that, at each moment of time, the Doppler factor within the optical radiating region has an approximately constant value. 
In the radio range, the emission detected by single-dish antennae from flat spectrum radio sources, which includes OJ~287, is mainly generated in a compact VLBI core (e.g., \cite{Kovalev05}).
The distance of the VLBI core observed at a given frequency from the true jet base is almost constant \cite{BK79, Lobanov98, Pushkarev12}, although changes are possible on average by 0.3~mas for typical offset of the core positions at 2 and 8~GHz of about 0.5~mas \cite{Plavin19}.
Hence, we supposed that the radio-emitting region is compact enough to have one value of the Doppler factor of all its parts and is located at a fixed distance from the cone apex. 
Then, the changes in $\delta$ caused by variations in the azimuth angle $\varphi$ of the region, from which the observed radiation at a certain frequency comes, should appear in the long-term light~curve.

The flux variability period produced by a change in $\delta$ depends on a rate of a change in $\varphi$ of components crossing through the emission region.
In other words, in their outward motion, the components of the helical jet pass consecutively first through the region in which such parameters are reached, under which optical radiation reaches the observer without a strong absorption.
After passing through this region, the magnetic field strength, electron density, and possibly energy spectrum of emitting particles become such that it is impossible to generate a significant flux density of synchrotron optical radiation.
Moving further down the jet, the physical parameters in the components change, and at a certain fixed distance from the true jet base, the component becomes less opacity for 15~GHz radiation, forming the 15~GHz VLBI core, and the component becomes optically thin for the 15~GHz radiation further downstream. 
As shown in \cite{But18b}, when components move along radial (ballistic) trajectories, the change rate of $\varphi$ for components that reach a certain fixed distance from the cone apex does not depend on the distance to the cone apex.
Then, the periods of optical and radio flux variability would have to coincide.
However, if the jet components move at the angle $p\neq0^\circ$ to the radial trajectory, then the further away from the cone apex, the lower the rate of change in $\varphi$ and, therefore, the longer the period of $\delta$ change, which is expressed as \cite{But18b}
\begin{equation}
   T_\text{var}=\frac{2 \pi r \sin\xi\cos \rho}{\beta\, c \sin \left( \rho-p\right)}, 
    \label{eq:Tvar}
\end{equation}
where $r$ is a distance from the cone apex to a cross-sectional plane passing through the region, in which the observed radiation at a given frequency is generated.

The motion of the VLBI jet features of the OJ~287 jet is detected by the VLBA monitoring programs at 15~GHz \cite{Lister19} and 43~GHz \cite{Agudo12}.
This suggests that $p>0^\circ$.
Note that, if $p=\rho$, the motion of components will occur along the jet flow. 
The jet will look stationary in time and space, and the periods of variability of both the flux density and the PA$_\text{in}$ will be absent.

\begin{figure}[H]
\centering
\includegraphics[width=15 cm]{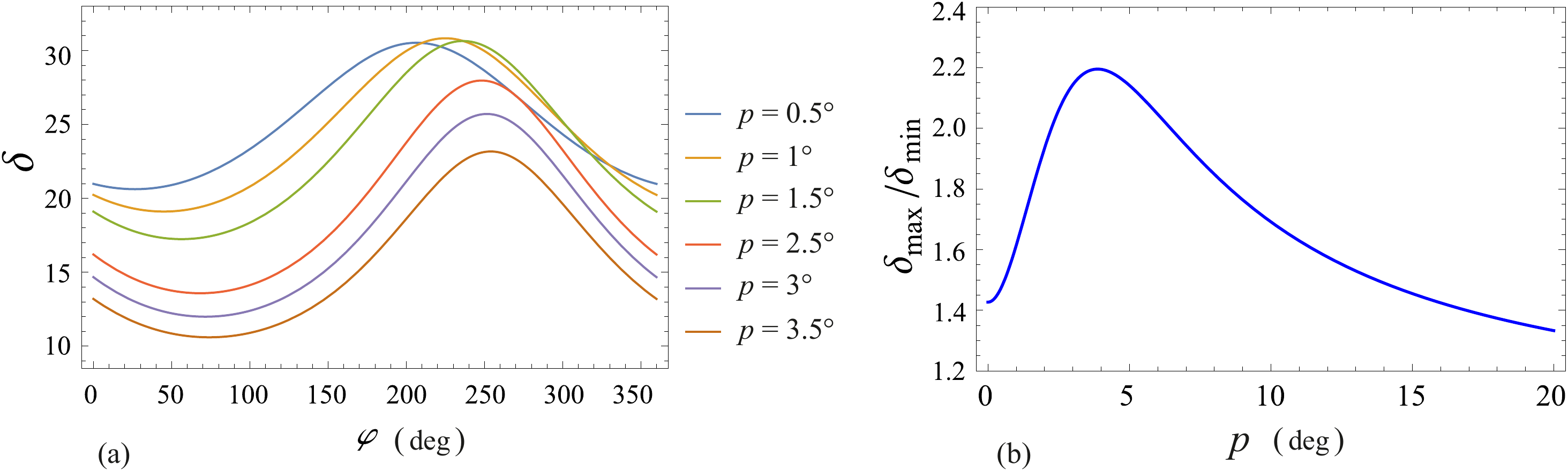}
\caption{Dependence of the Doppler factor on $p$: \textbf{(a)} Doppler factor versus the azimuth angle for various $p$ and \textbf{(b)} ratio of maximum and minimum values of Doppler factor as a function of $p$.
}
\label{fig:doppler}
\end{figure}

The non-radial motion of components leads to the variations of the Doppler factor of the emitting region caused by $\varphi$ variation being significantly stronger than under the ballistic motion of components \cite{But18a}.
Plot of $\delta\left( \theta_0,\xi,p,\varphi\right)$ versus $\varphi$ for the adopted above values $\theta_0=1.5^\circ$, $\xi=1^\circ$ and a few values of $p$ is displayed in Figure~\ref{fig:doppler}. 
For this, the expression for $\theta=\theta\left(\theta_0, \xi, p,\varphi\right)$ obtained in Appendix
\begin{equation}
   \cos \theta=\cos p \left(\cos \xi \cos \theta_0-\sin \xi \sin \theta_0 \cos \varphi \right)-\sin p \sin \theta_0 \sin \varphi 
    \label{eq:cosTheta}
\end{equation}
is substituted into Equation~(\ref{eq:doppler}).
Values of $\theta$ from Formula~(\ref{eq:cosTheta}) are equal to those calculated from 
Formulae~(11)--(13) in \cite{But18a} under the corresponding parameters. 
Figure~\ref{fig:doppler} shows that maximum values of $\delta$ are reached for small $p$.
For example, if $p<3^\circ$, then $\delta_\text{max}>25$, while at $p=5^\circ$, the maximum value is no higher than $\delta=16$ .
Since the 12-year optical flares have large amplitudes, for the interpretation of these flares by a change in $\delta$, it is natural to assume such $p$ at which the following conditions are fulfilled.
First, $\delta\left(\varphi,p\right)$ reaches a high value at the maximum. 
Second, the ratio of maximum and minimum values of $\delta$ is close to the maximum of possible ones. 
Figure~\ref{fig:doppler} shows that it occurs at $p=2.5^\circ$. 

\begin{figure}[H]
\centering
\includegraphics[width=7 cm]{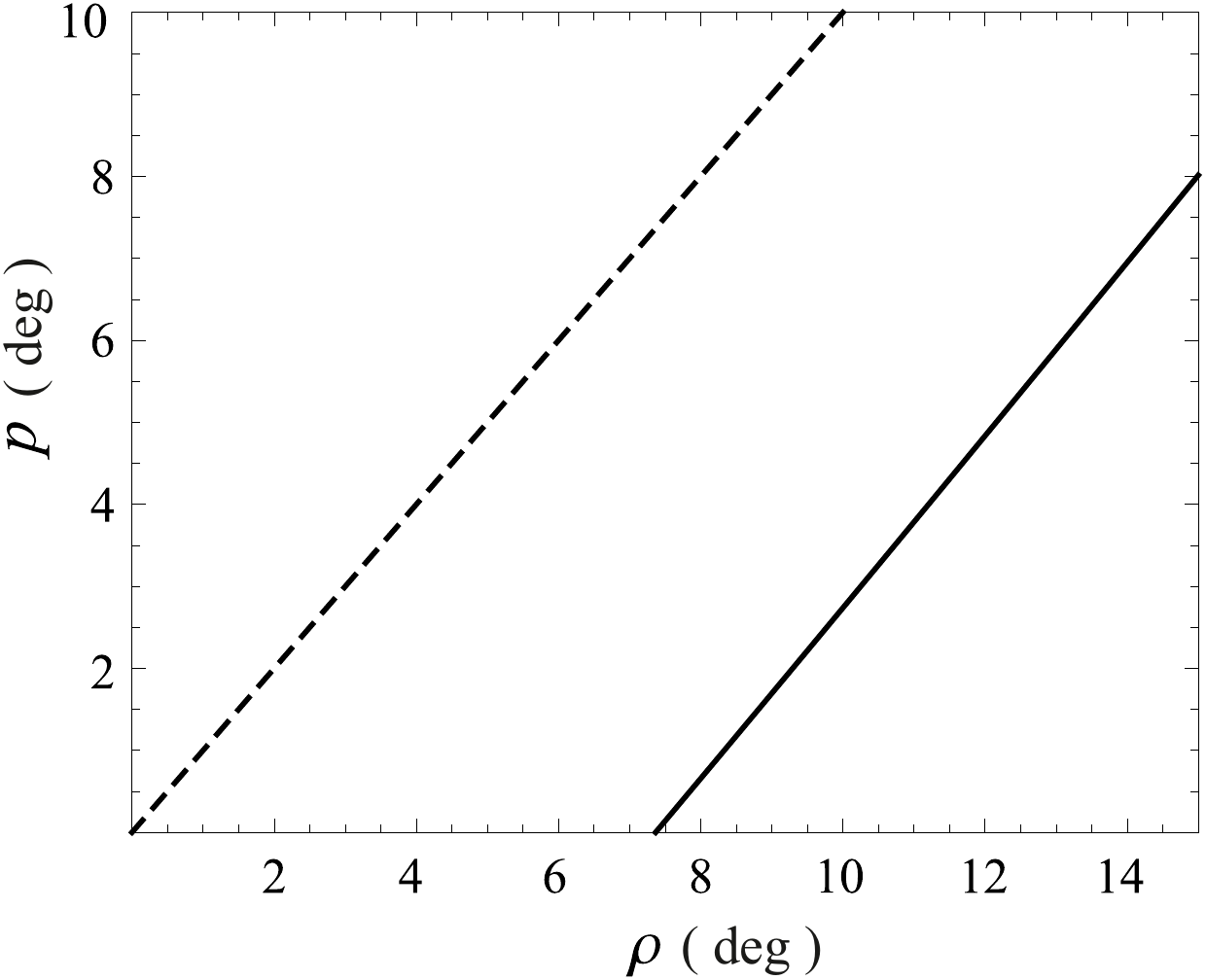}
\caption{The relation of the angles of the velocity vector and the jet flow local axis to the cone generatrix (solid line), given by Formula~(\ref{eq:Tvar}). The dashed line indicates $p=\rho$. 
}
\label{fig:pvsro}
\end{figure}

To use Formula~\ref{eq:Tvar} to find the angle $\rho$ between the tangent to the jet flow at a certain point and the cone generatrix drawn through this point, one needs to know the values of $T_\text{var}$ and $r$.
From the measurement of astrometric shift of the VLBI core observed at different frequencies, performed in \cite{Pushkarev12}, $r$ can be found from
\begin{equation}
    r=r_\text{core,\,15\,GHz} \cos \rho,
    \label{eq:r_coreshift}
\end{equation}
where $r_\text{core,\,15\,GHz}=\Omega_{r\nu}/\left(\nu \sin \theta \right)$ is a distance of 15.4~GHz VLBI core from the true jet base, $\Omega_{r\nu}$ is the core shift measure defined in \cite{Lobanov98}, $\nu$ is the observed frequency in GHz, and $\theta$ is the jet viewing angle, which corresponds to $\theta_0$ in our case.  
As reported in \cite{Pushkarev12}, OJ~287 has $\Omega_{r\nu}<4.18$~pc~GHz, which gives $r_\text{core\,15\,GHz}<10.3$~pc.
According to this, we will not be much mistaken if we take $T_\text{var}=T_\text{PA}$, since PA$_\text{in}$ was determined by the inner jet part within about 0.1~mas from the 15.4~GHz core (see Section~2.2).
The dependence $p\left(\rho\right)$ given by Formula (\ref{eq:Tvar}) is shown in Figure~\ref{fig:pvsro}.
It can be seen that $p>\rho$ and $p\approx10^\circ$ always for $p=2.5^\circ$.

Let us consider whether the formation of the helical jet with such geometric parameters is possible due to the development of the Kelvin--Helmholtz instability, which can be originated due to a velocity difference across the interface between the jet and surrounding medium.
According to \cite{Hardee82}, a helical wave mode can be represented as a wave propagating at some angle to the cone axis. This angle is 
\begin{equation}
\sin \rho=\frac{2\pi R}{\left[\lambda^2_\text{h, obs}+\left(2\pi R\right)^2\right]^{1/2}},
    \label{eq:sinRO}
\end{equation}
where $R$ is the jet radius.
Based on numerical simulations, Hardee \cite{Hardee82} obtained the empirical expression for the wavelength at which the growth rate of perturbation for the helical wave mode is~maximum:
\begin{equation}
\lambda_\text{h}=1.6\frac{M_j}{\eta^{1/3}}R,
    \label{eq:lambdaH}
\end{equation}
where $M_j$ is the Mach number in the jet and where $\eta$ is the ratio of densities of a jet and its surrounding medium.
The estimation of $\lambda_\text{h}$ by Formula~(\ref{eq:lambdaH}) gives an accuracy of about $10\%$ \cite{Hardee82}.   
According to \cite{Perucho12}, the wavelength of the helical wave in the observer's reference frame is
\begin{equation}
    \lambda_\text{h, obs}=\lambda_\text{h}\frac{\sin \theta_0}{\left(1-\beta_\text{w} \cos \theta_0 \right)},
    \label{eq:lambdaOBS}
\end{equation}
where $\beta_\text{w}=\beta \cos p$ is the speed of the helical wave along the cone axis. 

\begin{figure}[H]
\centering
\includegraphics[width=10 cm]{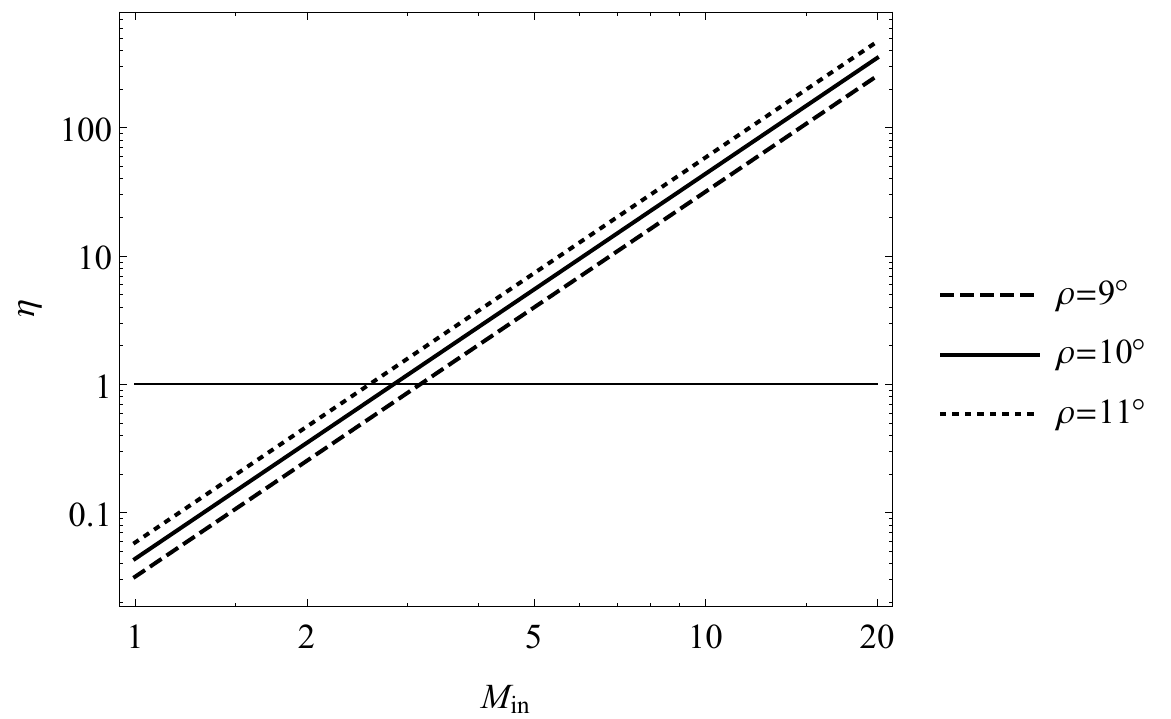}
\caption{Dependence of the ratio of the jet and surrounding medium densities $\eta$ on the Mach number in the jet flow for the parameters found for the blazar OJ~287 helical jet:
the horizontal line marks $\eta=1$.
}
\label{fig:KH}
\end{figure}

From Formulae~(\ref{eq:sinRO})--(\ref{eq:lambdaOBS}), we plotted the dependence $\eta\left( M_j\right)$ for the jet parameters found above (Figure~\ref{fig:KH}).
It is seen that, for creation of the helical jet, which is able to explain the observed period of radio variability, the sets of $M_j$ and $\eta$ parameters lie in the range of adequate values.

The difference in the periods of optical and radio (both the flux density and positional angle of the inner jet) variability can be caused by different distances between the jet origin and the regions responsible for the observed variability.
The ratio of these distances is equal to the ratio of the corresponding variability periods \cite{But18b}. 
Then, the optical emitting region is at the distance of
\begin{equation}
    r_\text{opt}=r_\text{15 GHz}/T_\text{PA}\approx 4.2\text{~pc}
    \label{eq:ropt}
\end{equation}
from the true jet base.
This distance appears to be very large, $\approx6\cdot10^4$ gravitational radii for a black hole of $10^9$ solar masses.
However, the estimation from Formula~(\ref{eq:ropt}) was made under the assumption that the jet axis lies on the cone surface.
In theoretical works \cite{BesNokh06, Beskin17}, it is shown that, near its origin, the jet propagates within the paraboloid and starting from a certain distance: within the cone.
Based on the VLBA observation for jets of nearby active galaxies, the transition from the parabolic to conical jet shape occurs at distances $10^5-10^6$ gravitational radii \cite{Pushkarev17, Kovalev20}.
Taking this into account will reduce $r_\text{opt}$, but it will not change the interpretation of the difference in variability periods based on a spatial displacement of the regions responsible for dominating optical and radio emission.
These jet parts are at different separations from the central machine and thus have different azimuth angle, causing changes in the Doppler factor and PA$_\text{in}$ and occurring with different rates under other unchanging geometrical and kinematic parameters of the model, listed in Table~\ref{tab:heljetpar}.

\section{Helical Jet Precession}

If the 12-year flares are formed only due to the increase in $\delta$, then, according to Formula~(\ref{eq:flux}), the maximum value of $\delta$ determines the flux peak value.
Inserting Formula~(\ref{eq:cosTheta}) into (\ref{eq:doppler}), we found that the function $\delta\left( \varphi\right)$ has extremum at
\begin{equation}
    \varphi_\text{extr}=\arctan\left(\frac{\tan p}{\sin\xi} \right)+\pi n,
    \label{eq:FIextr}
\end{equation}
For $n=0$ and $n=1$, the Doppler factor has minimum $\left(\delta_\text{min}\right)$ and maximum  $\left(\delta_\text{max}\right)$, respectively. 
Hence, for constants $p$ and $\xi$, the values of $\delta_\text{max}$ and $\delta_\text{min}$ are invariable and occur at a fixed $\varphi_\text{max}$ and $\varphi_\text{min}$, correspondingly. 
If the flux density in the reference frame of the jet is constant, the flux peaks in the observer's reference frame must be equal as well.

Since $\delta$ at the maximum of the flux density reaches a fixed value, the difference in peak fluxes of 12-year flares can be caused either by an internal change in the jet or by a change in the angle of the cone axis with the line of sight.
In the first case, the ratio of peak fluxes at the maximum Doppler factor (e.g., $\approx3$ for the flares of 1983 and 1995), 
would have to be maintained at other values of $\delta$, e.g., at $\delta_\text{min}$.
The observed minimum flux density $F_\text{min}$ between the flares differs only slightly \cite{Villforth10}.
Therefore, in order to agree with the observed long-term light curve, the increase in the flux density in the reference frame of the jet would have to occur at a time when the Doppler factor for an earth's observer would be close to the maximum value.
There is no physical ground for this condition.

In our view, the most possible explanation for the observed difference in the peak fluxes of the 12-year flares is a change in the angle $\theta_0$, which, for example, may be associated with precession.

\begin{figure}
\centering
\includegraphics[width=9cm]{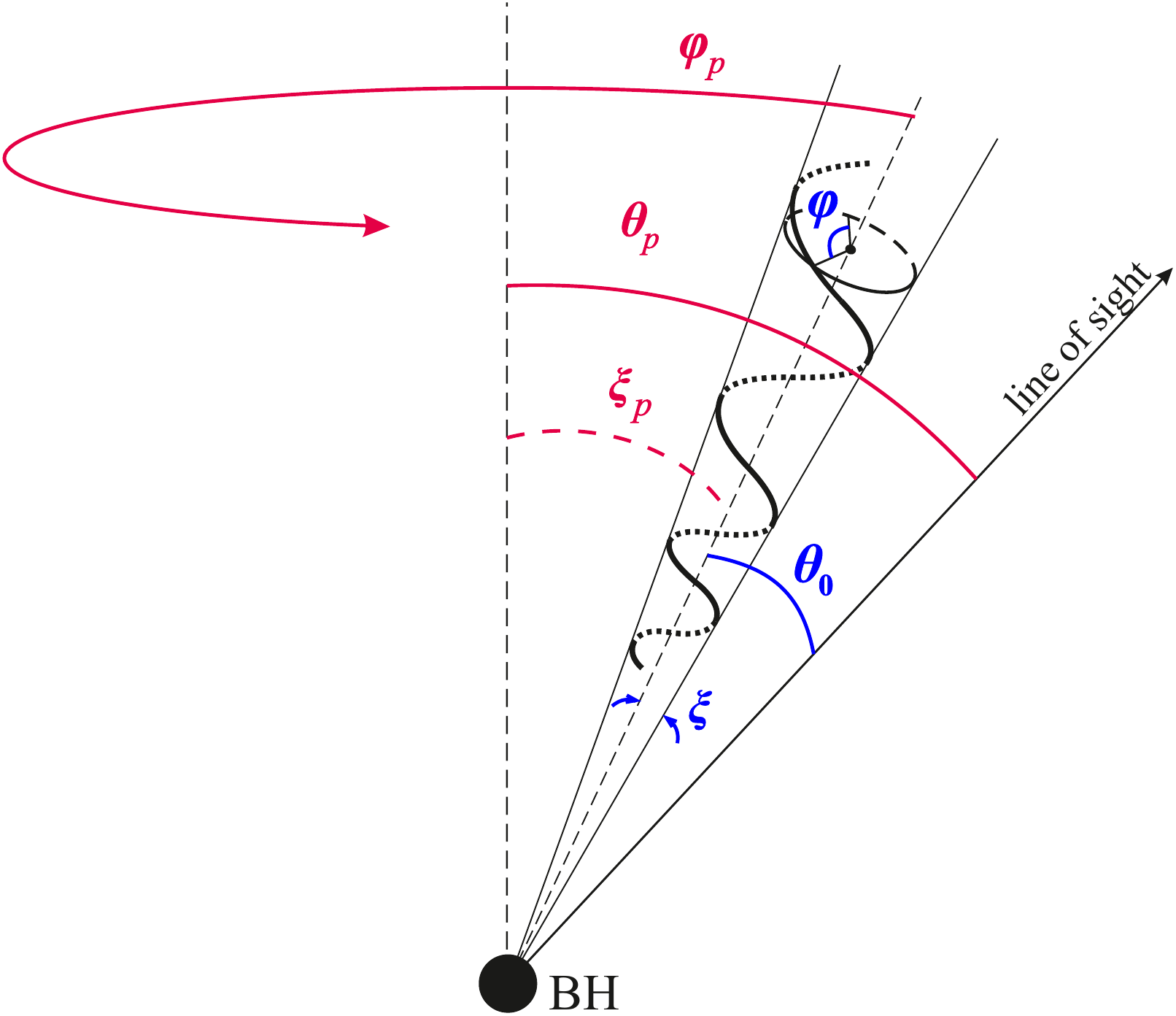}
\caption{A sketch of the helical jet precession:
the blue color marks geometrical parameters of the helical jet, which is originated by development of the Kelvin--Helmholtz instability.
The red color marks geometrical parameters of the precession of the helix axis under the assumption of  a single supermassive black hole being in the center of OJ~287.
}
\label{fig:theta_heljetprec}
\end{figure}

Let us check this assumption.
The scheme of the helical jet precession is presented in Figure~\ref{fig:theta_heljetprec}. 
The dependence of $\theta_0$ on the azimuth angle of precession $\varphi_p$ is expressed as    
\begin{equation}
    \cos \theta_0\left(\varphi_p \right)=\cos\xi_p \cos \theta_p-\sin \xi_p \sin \theta_p \cos \varphi_p ,
    \label{eq:costhetapr}
\end{equation}
where $\xi_p$ is the half-opening angle of the precession cone and $\theta_p$ is the angle of the precession axis with the line of sight.

\begin{figure}
\centering
\includegraphics[width=10 cm]{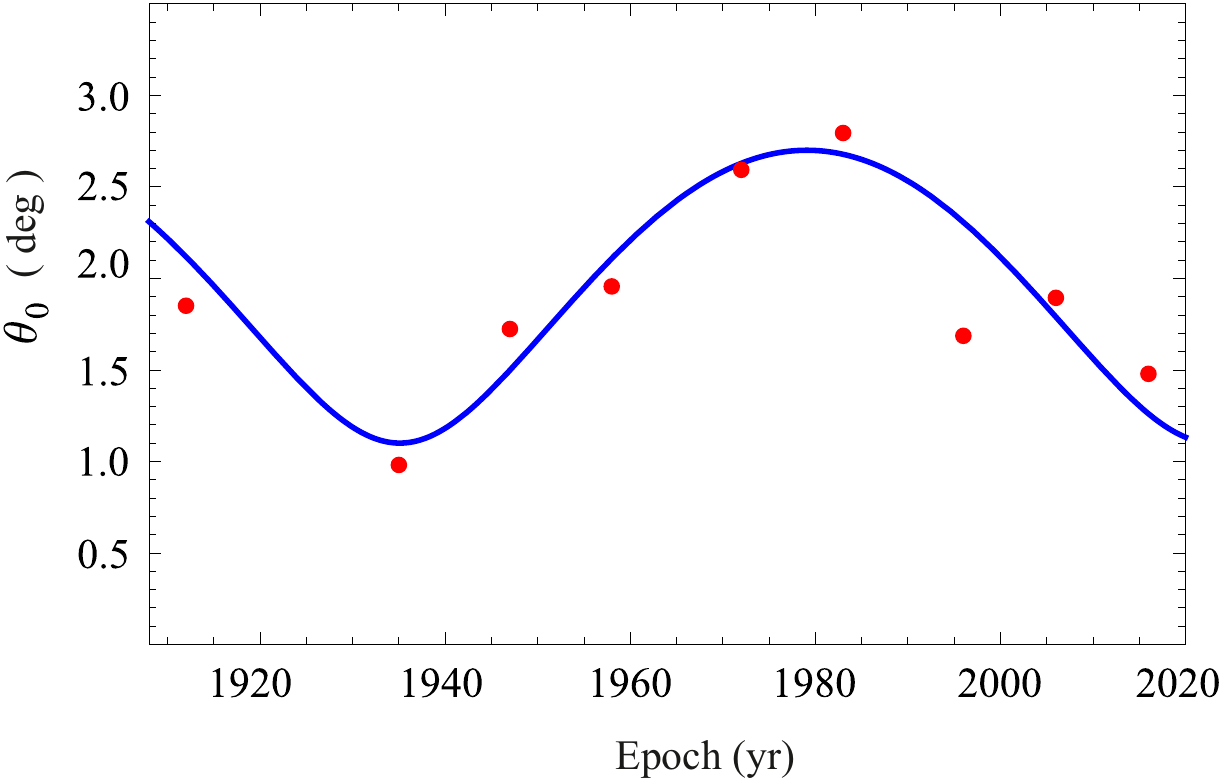}
\caption{Changes of the angle between the jet helix axis with the line of sight $\theta_0$ due to the precession of the OJ~287 jet. 
}
\label{fig:theta0prec}
\end{figure}

As it follows from Equation~(\ref{eq:FIextr}), the azimuth angle values, for which the Doppler factor achieves the maximum and minimum values which do not depend on $\theta_0$.
Assuming that the flux density in the reference frame of the emitting plasma is constant, we obtained from Equation~(\ref{eq:flux})
\begin{equation}
    \frac{\delta_\text{max}\left(\varphi_\text{max}, \theta_0,p,\xi \right)}{\delta_\text{min}\left(\varphi_\text{min}, \theta_0,p,\xi \right)}=\left(\frac{F_\text{max}}{F_\text{min}}\right)^{1/(2+\alpha)}.
    \label{eq:Doppler_ratio}
\end{equation}
Inserting into Equation~(\ref{eq:doppler}) the expression for $\theta$~(\ref{eq:cosTheta}), in which $\theta_0$ and $\varphi_\text{max, min}$ are given by Formulae~(\ref{eq:costhetapr}) and (\ref{eq:FIextr}), correspondingly, and solving Equation~(\ref{eq:Doppler_ratio}) with respect to $\theta_0$, substituting the observed flux densities of the 12-year peaks and minima between them, we obtained    
\begin{equation}
\theta_0=\arcsin \left(\frac{k_3}{\sqrt{k_1^2+k_2^2}} \right)-\arccos \left( \frac{k_1}{\sqrt{k_1^2+k_2^2}}\right),
    \label{eq:theta_0}
\end{equation}
where
$$k_1=\beta \left( \frac{\delta_\text{max}}{\delta_\text{min}}+1\right) \left(-\sin p \sin \varphi_\text{max}-\cos p \sin \xi \cos \varphi_\text{max} \right), $$
$$k_2=\beta \cos p \cos \xi \left(\frac{\delta_\text{max}}{\delta_\text{min}}-1 \right),$$
$$k_3=\left( \frac{\delta_\text{max}}{\delta_\text{min}}-1\right).$$
We have taken into account that $\varphi_\text{min}=\varphi_\text{max}-\pi$ for the range of azimuth angle from 0 to $2\pi$.
The parameters used in Formula~(\ref{eq:theta_0}) and derived viewing angle $\theta_0$ values for each peak are listed in Table~\ref{tab:parprec}.
Fitting the $\theta_0$ values by the expression~(\ref{eq:costhetapr}), we obtained $\theta_p=1.8^\circ$, $\xi_p=0.7^\circ$ and the precession period in the observer's reference frame $T_p=92\pm8$~years.
The dependence of $\theta_0$ with time is shown in Figure~\ref{fig:theta0prec}. 

\begin{table}[h]
\caption{Used observational data \cite{Villforth10, Valtonen19} and obtained values of the angle between the cone axis and the line of sight $\theta_0$: the columns contain (1) mean data of 12-year flares; (2) maximum flux density of the flare; (3) mean minimum flux density after the corresponding flare; and (4) obtained values for $\theta_0$. }
\label{tab:parprec}
\centering
\begin{tabular}{cccc}
\toprule
\textbf{year}	& \textbf{$F_\text{max}$, mJy}	& \textbf{$F_\text{min}$, mJy} & $\theta_0$, deg\\
(1) &   (2) &   (3) &   (4) \\
\midrule
1912		& 40			& 3     & 1.9\\
1932		& 13			& 3     & 1.0\\
1947        & 23            & 2     & 1.7\\
1958        & 30            &2      & 2.0\\
1972        & 55            & 2     & 2.6\\
1983        & 32            & 1     & 2.8\\
1996        & 11            & 1     & 1.7\\
2006        & 14            & 1     & 1.9\\
2016        & 17            & 2     & 1.5\\
\bottomrule
\end{tabular}
\end{table}

In order to find the precession period $T'_p$ in the reference frame of the source, it is necessary to take into account not only the cosmological time change but also that the time interval in the source reference frame is $t'=\delta t$. 
Since $\delta$ changes over a wide range (see Figure~\ref{fig:doppler}), we have divided the whole precession period in the source reference frame into equal time intervals.
Assuming that the changes of $\varphi_p$ in the jet reference frame occur uniformly and $\varphi_p$ changes are 7 times faster than $\varphi$, we took the time for which $\varphi_p$ changes by $1^\circ$ as a time interval unit.
Then,
\begin{equation}
T^\prime_p=\frac{T_p}{\left(1+z \right)\sum_{i=1}^{360}\frac{1}{360 \delta_i}},
    \label{eq:Tprec}
\end{equation}
where $\delta_i$ is the Doppler factor at the end of the time interval.
From~(\ref{eq:Tprec}), we obtained $T'_p\approx1.2\cdot10^3$~years.

\begin{figure}[H]
    \centering
    \includegraphics[width=10 cm]{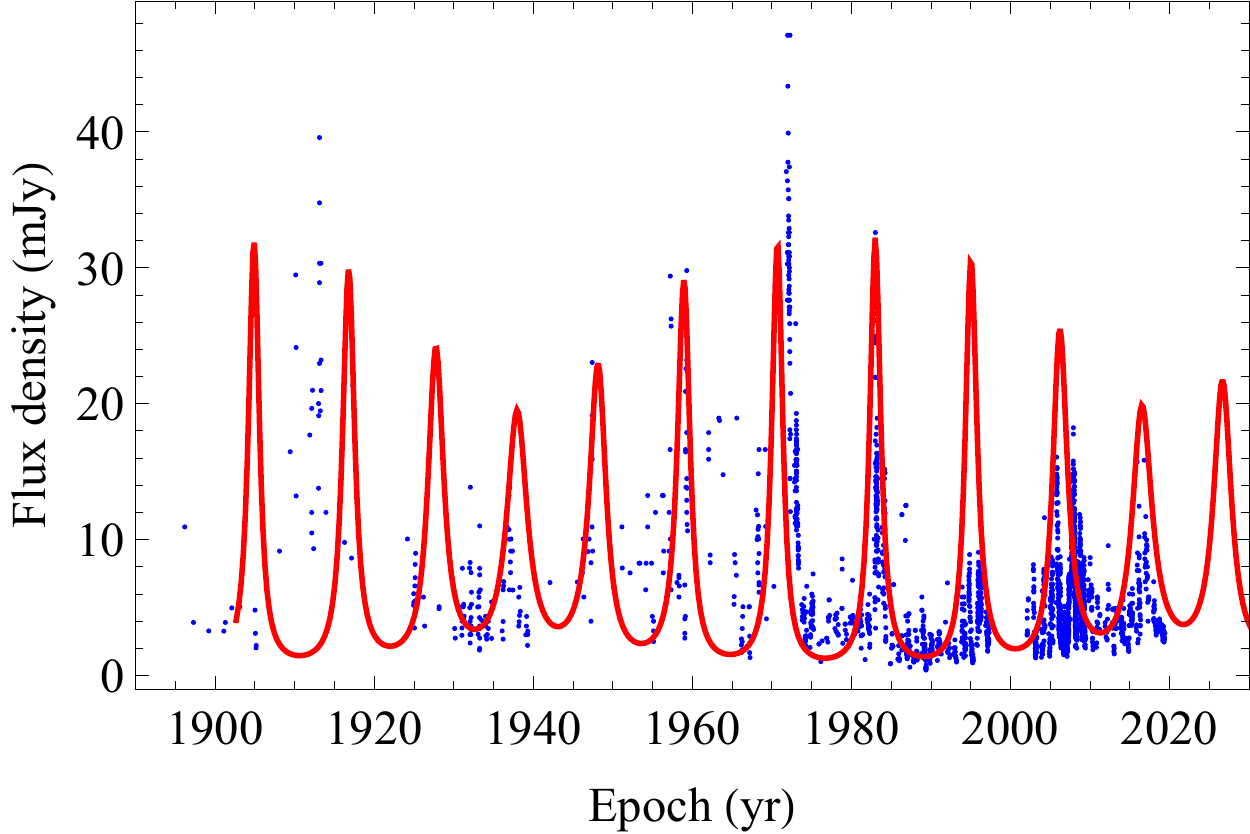}
    \caption{Observed data points (blue circles) and simulated light curve (red line) in V band.}
    \label{fig:LC}
\end{figure}

Thus, by substituting $\theta_0$ from (\ref{eq:costhetapr}) into  Formula~(\ref{eq:cosTheta}), we can obtain an analytical expression to describe the change in the Doppler factor of the radiating region in the jet. 
Assuming that the 12-year flares are caused by a change in $\delta$, we modeled the light curve. 
For this aim, we took the 1983 flare as a reference point. 
From Formula~(\ref{eq:FIextr}), we found the azimuth angle of the optical radiating region $\varphi=248.2^\circ$. 
Using it together the data from Table~\ref{tab:parprec}, we found the azimuth angle for the precession motion $\varphi_{p}=15.4^\circ$ by Formula~(\ref{eq:costhetapr}). 
The Doppler factor at a given time moment depends on $\varphi$ and $\varphi_{p}$. 
The rate of $\varphi_{p}$ change was determined from $T^\prime_{p}$, and it is approximately $1.5^\circ$ for 5 years in the source reference frame. 
Figure~\ref{fig:theta0prec} shows that $\varphi$ changes about 7 times faster than $\varphi_{p}$. 
The flux density was found by Formula~(\ref{eq:flux}) and normalized to the flux density of the 1983 flare.
Figure~\ref{fig:LC} shows the simulated light curve and observational data from the literature and the archive of the Crimean Astrophysical Observatory. 
As seen, since 1940, the observational data are in a good agreement with the simulated flux density. 
Some difference between the observed and modeled flux densities at the maxima may be caused by a small physical change in the flux density in the source reference frame compared to that of the 1983 flare maximum. 
Here, we do not model the radio light curve because it has no prominent maxima as in the optical range. 
There is no reliable reference point for which the values $\varphi$ and $\varphi_{p}$ can be confidently found. 
Since the optical and radio radiating regions are spatially separated, the azimuth angles of the radiating regions are different within the framework of the considered model.

\begin{figure}[H]
\centering
\includegraphics[width=10 cm]{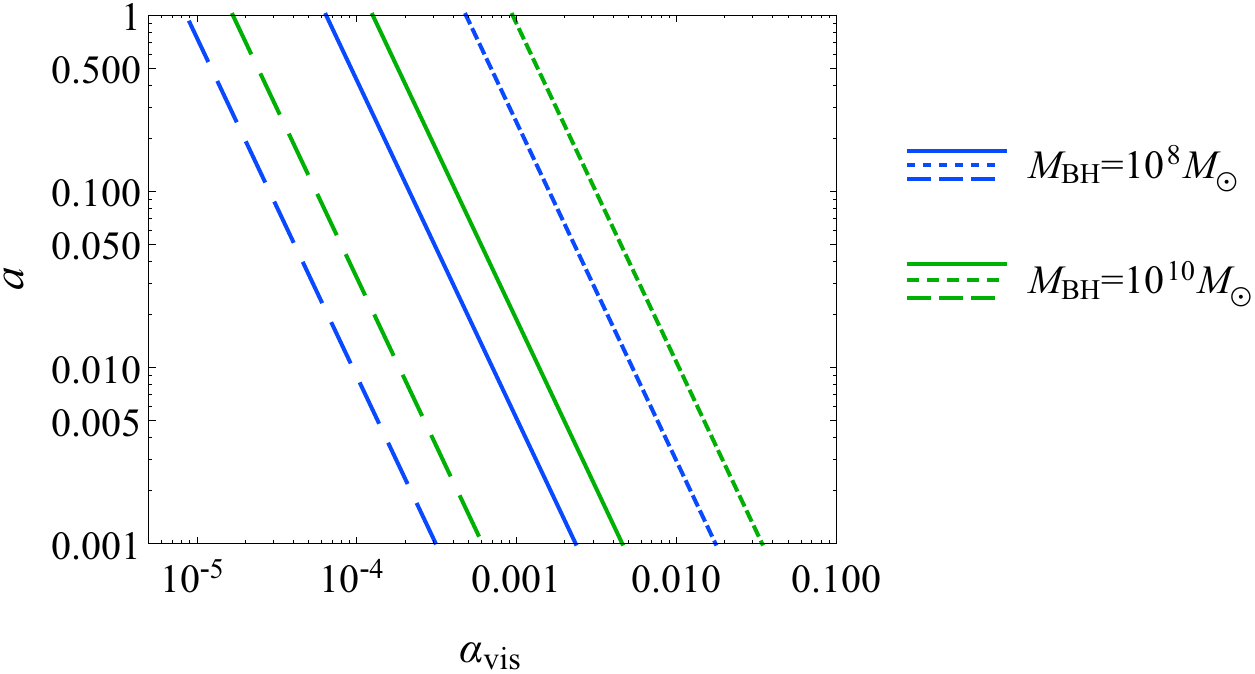}
\caption{Relations of parameters $a_\text{BH}$ and $\alpha_\text{vis}$ for various $M_\text{BH}$ and $\dot{m}$ for period $T'_p\approx1.2\cdot10^3$ years of the precession of both the black hole and the inner part of the accretion disc: dashed, solid, and dotted lines correspond to $\dot{m}$ equal to 0.001, 0.01, and 0.1 in Eddington units, respectively.
}
\label{fig:LTBH}
\end{figure}

Let us check whether the Lense--Thirring precession with such a period can occur in the accretion system of the single supermassive black hole. In this system, the precession of the black hole together with the inner part of the accretion disc can be originated in result of misalignment of the black hole and accretion disc outer part rotation axes, as described in \cite{Lu92}. This precession  period can be expressed as
\begin{equation}
    T'_p=10^{9.25}a_\text{BH}^{5/7}\alpha_\text{vis}^{48/35}\left(\frac{M_\text{BH}}{10^8 M_\odot} \right) \left(\frac{\dot{M}_\text{BH}}{10^{-2}M_\odot \text{ yr}^{-1}} \right)^{-6/5} \text{ yr},
    \label{eq:TLTBH}
\end{equation}
where $a_\text{BH}$ is the dimensionless parameter expressing a black hole's specific angular momentum, $\alpha_\text{vis}$ is the viscosity parameter, and $M_\text{BH}$ and $\dot{M}_\text{BH}$ are the mass and mass accretion rate of the black hole, respectively.
We expressed the Formula~(\ref{eq:TLTBH}) as
\begin{equation}
    T'_p=10^{6.43} a_\text{BH}^{5/7} \alpha_\text{vis}^{48/35} \dot{m}^{-6/5} \left(\frac{M_\text{BH}}{10^8 M_\odot} \right)^{-1/5}  \text{ yr},
    \label{eq:TLTBHedd}
\end{equation}
where $\dot{m}$ is the mass accretion rate in Eddington units.
Using the obtained period $T'_p$ in Formula~(\ref{eq:TLTBHedd}), we plotted a dependence of $a_\text{BH}\left( \alpha_\text{vis}\right)$ for different masses and mass accretion rates of the black hole (Figure~\ref{fig:LTBH}).
The range of possible values for the black hole mass is large enough. According to the broad-band observations, black hole masses for BL Lacertae objects were estimated as $\sim (10^8-10^9) M_\odot$ \cite[][and references therein]{SbarratoGhM12,TitarchukSeifina17}. Valtonen et al. \cite{Valtonen19} derived the primary black hole mass of $\sim 10^{10} M_\odot$ under the assumption of the binary black hole system in the OJ~287 centre. This mass estimation is model-dependent, but we took it as an upper limit.
The mass accretion rate we accepted is in the range $\dot{m}=0.001-0.1$, according to the values reported in \cite{Xie04, XiongZhang14}. 
As seen, the dependence on the black hole mass is weak.
The precession of a slowly rotating black hole with a high mass accretion rate and a low $\alpha_\text{vis}$ (as confirmed by observations \cite{King07}) can occur with the found period. 

On the other hand, it is possible that misalignment of the black hole and accretion disc rotation axes can cause precession of the outer part of the accretion disc with a period \cite{Bhatta18}:
\begin{equation}
   T'_{p}=0.18\left(\frac{1}{a_\text{BH}} \right) \left(\frac{M}{10^9 M_\odot} \right) \left(\frac{r}{r_g} \right)^3 \text{ days},
    \label{eq:Tprecdisc}
\end{equation}
where $r$ is a radial distance of the accretion disc precessing part from the black hole and $r_g$ is the gravitational radius of the black hole.
From Formula~(\ref{eq:Tprecdisc}), it follows that, for $a_\text{BH}=0.01-1$, the regions of the accretion disc precessing with the found period are located at distances of $(30-300)r_g$ from the black hole with $M_\text{BH}=(10^8-10^{10})M_\odot$.
These distances seem to be reasonable to form the jet by the Blandford--Znajek mechanism \cite{BZ77}.

Note that, here, we investigate the possibility of the formation of the 1200 year period in the accreting system of a single black hole. More precise determination of the parameters of this system, such as the mass, mass accretion rate, and viscosity, should be based on the broad-band photometric and VLBI data.

\section{Discussion and Conclusions}

OJ~287 is the first blazar for which  periodicity in the optical light curve was detected, revealing prominent flares that occur approximately every 12 years \cite{Sillanpaa88}.
At that time, the unified scheme of active galactic nuclei \cite{UP95} had not yet been proposed.
Therefore, early investigations of OJ~287 were predominantly focused on the interpretation of optical periodicity.
It was supposed that such a short variability period can be formed in a binary black hole system \cite{Sillanpaa88}.
The secondary component moves along its elliptical orbit and passes the pericenter every 9 years (in the rest frame), invoking a tidal effect on the accretion disc of the primary black hole.
As a result, the matter of the accretion disc flows into the black holes, leading to the observed flare on the light curve \cite{Sillanpaa88}.
This model does not take into account the structure of the 12-year flares, namely, that there are two peaks in each flare.
The interpretation of this fact given in \cite{LehtoValtonen96} is that the secondary black hole passes through the accretion disc of the primary one twice during the orbital period.
This model also provided the explanation for some of the observed properties of optical polarization, but it was focused on orbital motion.
After the next 12-year flare occurred during 1995--1996, this model has been refined \cite{Valtonen06, Valtaoja00, ValtonenPihajoki13}.
The authors of \cite{ValtonenPihajoki13} predicted future evolution of the VLBI jet position angle observed at 15 and 43~GHz, but it has not been confirmed by later observations \cite[here and][]{Agudo12, Cohen17}.

An alternative interpretation of the presence of two peaks in a 12-year flare was proposed~in~\cite{VR98} considering two supermassive black holes of the same mass, with accretion discs and jets.
These jets have a helical shape and are twisted together.
Doppler beaming of a part of each of two jets, responsible for optical emission, can lead to the 2-peaked 12-year flares.

As it is noted in \cite{Sillanpaa88}, due to the radiation of gravitational waves, a separation between the pericenters of the primary and secondary black holes is expected to decrease, leading to an increase in the intensity of next 12-year flares.
The flares observed in 1995--1997 and 2005--2007, on the contrary, had significantly less maximum flux density ($\approx10-15$mJy in V band) than the previous flares in 1973 and 1984 (55 and 30~mJy in V band, respectively) \cite{Villforth10}.
Interpretation of the decrease in the maximum flux density of the 12-year flares as well as prediction of the absence of the flares in the future is proposed in \cite{Villforth10}.
These are explained by  something, that occurred before the $\approx$1970 event, causing an avalanche-like accretion of the disc magnetic field, which repeats every $\approx$12~years \cite{Villforth10}.
After each act of this accretion, the poloidal magnetic field near the black hole becomes stronger and more strongly hinders accretion of the poloidal magnetic field next to an act of accretion.
This model requires further verification by magnetohydrodynamic simulations.
The maximum flux density of the last 12-year flare occurred in 2016 was 17~mJy in B~band \cite{Valtonen19}.
It is slightly higher than the ones for two previous flares (Table~\ref{tab:parprec}). 

None of the above models can consistently explain the main observable facts: 
(i) the sudden change in PA$_\text{in}$ which was first detected at 43~GHz in 2004 \cite{Agudo12, Cohen17} and about six years later at 15~GHz; 
(ii) the difference of the long-term variability period in optical ($\approx$12~yr \cite{Sillanpaa88}) and radio ($\approx$25~yr \cite{Ryabov16,Donskykh16,Sukharev19}) flux densities.

In this paper, we explain most of the observed long-term properties of the blazar OJ~287, while other properties can also be interpreted within the framework of the proposed model.
First of all, we assume that the OJ~287 jet has a helical shape because this shape can be naturally originated by the jet nozzle precession or development of hydrodynamic instabilities.
For the certain relation between the angle of the helix axis to the line of sight and the half-opening angle of the cone, the sudden change in the inner jet position angle occurs under the outward motion of the helical jet $\left( \theta_0/\xi \gtrsim 1 \right)$.
As it was shown in \cite{But18a}, for the permanent geometrical parameters of the jet helix, the sudden change in PA$_\text{in}$ occurs at a certain fixed azimuth angle. 
For the case of OJ~287, PA$_\text{in}$ has a minimum and maximum value at $\varphi$ of about $132^\circ$ and $228^\circ$, correspondingly.
Therefore, the fact that the sudden change in PA$_\text{in}$ occurred at a higher frequency first and then at a lower one indicates that the 43~GHz emitting region reached this value of $\varphi$ earlier than the 15~GHz emitting region.
This means that the regions responsible for dominant radiation at different frequencies are spatially separated. 
This fact follows from the jet model \cite{BK79} and has been confirmed by VLBI observations for a large number of AGN jets \cite{Pushkarev12}.
The higher the observed frequency, the closer to the true jet base is the location of the region, in which the observed radiation is generated.
Hence, the region from which the observed optical radiation comes is located much closer to the true jet base. 
According to this, the changes in the flux density in the optical and radio ranges, caused by the geometric effects, do not coincide in time, since the Doppler factor of the emitting regions differs.
The absence of sharp and prominent flares in radio similar to those present in optical range can be explained by a significantly larger size of the radio-emitting region compared to that in optics.
This difference in the size of the regions can naturally be caused by different cooling time scales of ultrarelativistic electrons producing synchrotron radiation in the radio and optical ranges.

Within the framework of the described jet model, the difference between the periods of long-term optical and radio variability occurs when the jet components move along non-radial trajectories.
Such motion is observed \cite{Agudo12, Cohen17} and most simply explained by the development of hydrodynamic instabilities in the jet, for example, Kelvin--Helmholtz instability \cite{Hardee82}.
Moreover, we have shown that the derived parameters of the helical jet can also explain the observed variability periods. 
These parameters are attributed to the jet, along which the helical mode of the Kelvin--Helmholtz instability is developed \cite{Hardee82}.

In the framework of our model, the Doppler factor changes with distance from the true jet base, and for each jet component, it varies with time. Thus, for each jet component, the equal time intervals in the observer’s reference frame are different. Also equal time intervals in the observer’s reference frame for simultaneously observed jet features are different. These facts lead to different apparent speeds for different jet features, as it is observed for many AGN jets including OJ~287 \cite{Lister19}.

With the geometrical and kinematic parameters of the OJ 287 jet unchanged for decades and under the interpretation of the 12-year optical flares by geometry effects, the difference in peak fluxes in these flares may be due to the fact that the helical jet axis precesses.
This precession in the our geometrical model of the observed optical flux density variations can additionally explain different flux densities between the different 12-year flares and different time interval in the observer's reference frame between the 12-year flares.
We found that the jet precession period in the source reference frame is $\sim$1200~years and showed that jet precession can be formed by the Lense--Thirring precession of a single supermassive slowly rotating black hole with a high accretion rate and a small viscosity parameter or by precession of the accretion disc at distances of $60$--$300$ gravitational radii from a single supermassive black hole.
The observed properties of OJ~287 are explained in the assumption of a single supermassive black hole in the center of the active nucleus, an important result for interpretaion of the blazar long-term flux periodicity.
The latter one is often explained by the binary black hole system in AGN, by analogy with OJ~287. 

The interpretation of the double-peaked structure of the 12-year optical flares is beyond the scope of this article.
However, the discussed above possible explanations for this observed fact do not contradict the proposed model.
Namely, the double-peaked flares can occur if some parts of the jet move at a small angle to the general helical trajectory \cite{But20},
or there are two jet regions radiating in the optical range, located at slightly different distances from the true jet base.
The Doppler factor of the more distant region changes as well as the Doppler factor of the region closest to the true jet base, only with a certain time delay.
Indications for the existence of extended optical jets have been recently found \cite{Kovalev17,PlavinKovPet19,KovalevZobnina20}.
With our model of the helical jet, we can make the following predictions for the blazar OJ~287: 
(i) an increase in the maximum flux density in the next three-four 12-year optical flares;
(ii) a gradual decrease in the position angle of the observed at 15~GHz inner jet until $\approx2038$, when again there will be a sudden increase in the position angle (it will happen a few years earlier at 43~GHz).

\authorcontributions{
M.S.B. have proposed methodology and have made conceptualization, and formal analysis; A.B.P. have curated the data; M.S.B. and A.B.P. have written original draft and revised it.
}


\funding{Theoretical part of this research was funded by Russian Science Foundation grant number 19-72-00105.}

\conflictsofinterest{The authors declare no conflict of interest.} 

\appendix
\section{The Angle of the Jet Component Velocity Vector to the Line of Sight }

\begin{figure}
\centering
\includegraphics[width=10 cm]{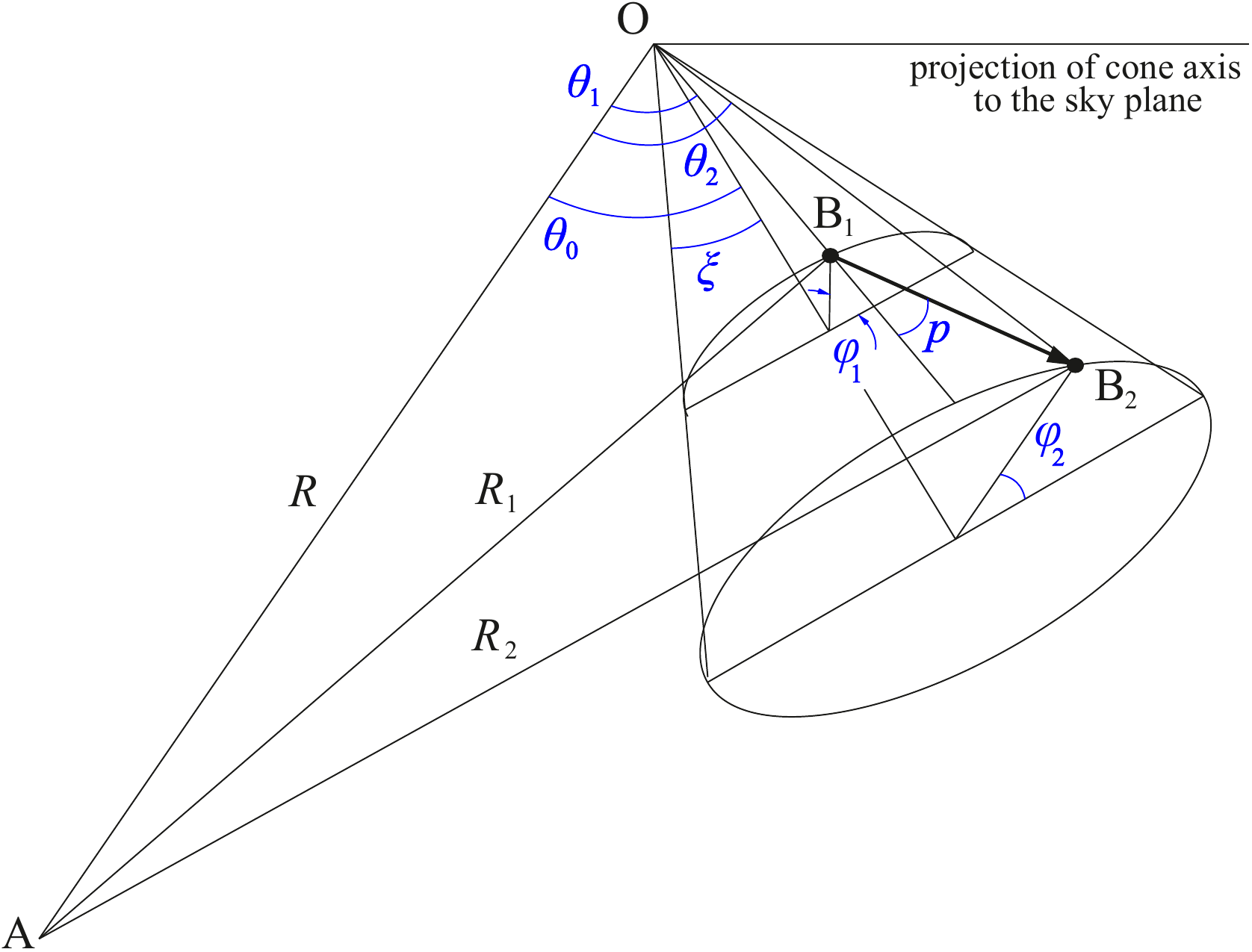}
\caption{Scheme of a jet component motion relative to the line of sight.
}
\label{fig:motion}
\end{figure}

We assume that the jet forms a helix on a surface of a notional cone.
Find the angle $\theta$ between the velocity vector \mbox{\boldmath $\upsilon$} of the jet component and the line of sight. 
In Figure~\ref{fig:motion}, point B$_1$ denotes a position of the jet component in some moment of time. 
Moving at the angle $p$ to the cone generatrix, the component is located at the point B$_2$ after a time interval $\Delta t$.
The observer is at point A.
From $\Delta$AB$_1$B$_2$, we have
\begin{equation}
   R_2^2=R_1^2+s^2-2R_1 s \cos \theta\approx R_1^2-2R_1 s\cos\theta, 
    \label{eq:tcos}
\end{equation}
where $R_1=$AB$_1$, $R_2=$AB$_2$, $s=$B$_1$B$_2=\upsilon \Delta t$. 
Approximate equality is obtained for $s\ll R_1$.
From Formula~\ref{eq:tcos}, it follows that
\begin{equation}
    \cos \theta \approx \frac{R_1-R_2}{s}.
    \label{eq:costheta}    
\end{equation}
$R_1$ can be found from $\Delta$OAB$_1$ taking into account that $d\ll R$
\begin{equation}
    R_1=\sqrt{R^2+d^2-2Rd\cos \theta_1}\approx\sqrt{R^2\left(1-\frac{2d}{R}\cos \theta_1 \right)}\approx R-d\cos\theta_1,
    \label{eq:R1}
\end{equation}
where $R=\text{OA}$, $d=\text{OB}_1$ is a distance of the component at the initial moment of time from the cone apex and where $\theta_1$ is the angle between the line of sight and the cone generatrix drawn through B$_1$. By analogy, from $\Delta$OAB$_2$, we have
\begin{equation}
R_2=R-\left(d+s\cos p \right)\cos \theta_2,
    \label{eq:R2}
\end{equation}
where $\theta_2$ is the angle between the line of sight and the cone generatrix drawn through B$_2$.  
The expression for $\theta_{1\,\left(2\right)}$  was derived in \cite{But18a}
\begin{equation}
    \theta_{1\,\left(2\right)}=\cos \xi \cos \theta_0 - \sin \xi \sin \theta_0 \cos \varphi_{1\, \left(2\right)},
    \label{eq:theta12}
\end{equation}
where $\varphi_{1\,\left( 2 \right)}$ is the azimuth angle for point B$_1$ $\left(\text{B}_2\right)$.
Inserting Formulae~(\ref{eq:R1})--(\ref{eq:theta12}) in (\ref{eq:costheta}) and taking into account $\varphi_2=\varphi_1-\Delta \varphi$ and $\Delta \varphi \ll 1$, we have
\begin{equation}
    \cos \theta = -\left( \frac{d}{s}+\cos p \right) \Delta \varphi \sin \xi \sin \theta_0 \sin \varphi_1+\cos p \left(\cos \xi \cos \theta_0-\sin \xi \sin \theta_0 \cos \varphi_1 \right).
    \label{eq:costhetamidle}
\end{equation}
In our case, $s \ll d$, $p \ll 1$, and $\Delta \varphi=s \sin p / \left(d \sin \xi \right)$ \cite{But18a}; the expression for $\theta$ is 
\begin{equation}
    \cos \theta=\cos p \left(\cos \xi \cos \theta_0 - \sin \xi \sin \theta_0 \cos \varphi_1 \right)-\sin p \sin \theta_0 \sin \varphi_1.
    \label{eq:costhetalast}
\end{equation}
 Note that if the jet twists in the opposite direction, i.e., $\varphi_2=\varphi_1+\Delta \varphi$, then  the sign of the last term in Formula~\ref{eq:costhetalast} changes.

\reftitle{References}


\vspace*{2mm}
{\bfseries Publisher's Note:} MDPI stays neutral with regard to jurisdictional claims in published maps and institutional affiliations.

\end{document}